\documentclass[a4paper,12pt]{article}
\usepackage{epsfig} % SRC Specials

\def\bea{\begin{eqnarray}}  \def\eea{\end{eqnarray}}
\def\1{{\rm 1\mskip-4.5mu l} }
\begin{document}

\title{\Large\bf\ Timelike Nonseparability and Retrocausation}

\author{{\large\bf Olivier Costa de Beauregard} \\
Fondation Louis de Broglie\\ 23 Quai de Conti, F-75006 Paris,
France}

%\date{ }

\maketitle

\vspace{2cm}
\begin{abstract}

Renewed interest in the quantum zigzagging causality model is
highlighted by an ingenious proposal by Suarez to test the timelike
aspect of nonseparability. Taking advantage of a work by
Fr\"{o}hner I argue that $\Psi$, the Dirac representation of a
state, has the Bayesian-like connotation of best estimate given the
Hilbert frame chosen. As a measurement perturbs uncontrollably a
system it is (Hoekzema's wording) a {\em retroparation}. My bet is
that Suarez' sources and sinks of paired particles operating inside
the coherence length of the laser beam will evidence
retrocausation.
\\

{\em Keywords:} zigzagging causality, measurement as retroparation,
$\Psi$ as best estimate given the Hilbert frame chosen, ``veiled
reality''. \\

\end{abstract}

\newpage
\pagestyle{plain}

{\bf 1}. Whenever chance occurrences show up as spacetime events,
correlated probabilities imply (via Bayes reversible formula)
reciprocally telegraphed information. Relativistic covariance is
then exigible.\par

Fourier analysis steps in; Fr\"{o}hner \cite{fr98}, in an
interesting paper, argues that it is inherent in the probability
formalism.\par

For the chance events handled by statistical mechanics the phase
cell is crucial, and its magnitude has physical meaning. De
Broglie, updating a deep remark of Hamilton and others (Klein,1890;
Vessiot,1906), likened action to phase via Planck's constant.
Heisenberg, Fourier-associating position and momentum, quantified
the phase cell. Finally Born showed that probability is a keystone
in the transition from geometric to wave mechanics. Fr\"{o}hner's
guided tour of the story uses Hamilton's equations in place of the
Hamilton-Jacobi extremum law.\par

Quantum mechanics is thus viewed as being jointly the physical
statistical mechan-ics and the physical telegraphic network. The
Bayesian reversal of joint probabilities entails action-reaction
reciprocity for spacelike separated events and cause-effect
reversibility for timelike separated ones. Retrocausation is thus
legalized.\par

Information pertaining to veiled reality is telegraphed via de
Broglie waves. The $\Psi$, Dirac's representation of a state, has
the Bayesian connotation of estimate of the state given the Hilbert
reference frame chosen -the one fitting the {\em preparing} or {\em
retroparing} \cite{dh92} device parading as a macroscopic
``object''.\par

So Dirac's {\em bra} or {\em ket} expresses a telegraphable
information (transition amplitudes including the propagator
\cite{oc83}). The much commented upon ``sum-and-product rule of
amp-litudes'' stems from a limitation in the telegraph.
Fr\"{o}hner, and before him Barut \cite{ab94}, assuming hiddenness
of a prepared EPRB spin zero state, derive classically the
correlation formula. How is this possible? Fr\"{o}hner answers:
``Quantum mechanics looks like an error propagating formalism for
uncertainty afflicted systems obeying the classical equations of
motion''.\par

Experimentation invalidates one EPR \cite{ae35} or Bell \cite{jb65}
assumption: A quantum measurement or observation unveil's not a
preexisting magnitude but, perturbing the system, contributes in
its realization. Wheeler's \cite{jw83} ``delayed choice
experiments'' stress this; for instance, ``measuring an electron
spin'' implies {\em first} aligning it arbitrarily, {\em second}
retroparing it as $\pm\frac{1}{2} h$. All this was part of the
twenties refrain, and should not have been forgotten.\\

{\bf 2}. A measurement thus is a reversed preparation -a {\em
retroparation} \cite{dh92}. Such is the basis of the zigzagging
causality model of EPR correlations \cite{oc53}. As the correlated
measured magnitudes do not preexist in the prepared state,
retrocausation is implied by definition. The EPR phenomenology
pinpoints the faulty EPR assumption: not identifying measurement as
retroparation.\par

The concise derivation \cite{oc83} of the EPRB formula thus allowed
is valid for both \cite{oc92} space or time distant occurrences.
Sutherland \cite{oc53} (a member of the zigzagging causality club)
temporarily resigned in 1985, frightened as he was by the
radicality of {\em retrocausation} in the timelike case; now
Tapster et {\em al} \cite{ptjr94} have evidenced it.\\

{\bf 3}. This author deems fundamentally deficient any analysis of
``quantum philosophy'' that is not explicitly {\em Lorentz-and-CPT
invariant}; he suspects it to conceal important phenomena. So he
bets that Suarez \cite{as98} very ingenious proposed test of
timelike non-separability will vindicate the zigzagging causality
model; he feels that the sources and sinks of photons operating
inside the coherence length of the laser beam will exhibit timelike
nonseparability.

\end{document}